%% file: Single_RF_Reflect.tex
\documentclass{article}
\usepackage{spconf,amsmath,graphicx}
\usepackage{amsmath,dsfont,bbm,epsfig,amssymb,amsfonts,amstext,verbatim,amsopn,cite,subfigure,multirow,multicol,lipsum,xfrac}
\usepackage{amsthm}
\usepackage{mathtools,amsthm}
\usepackage{perpage}
\usepackage{balance}
\usepackage{url}
\usepackage{amsfonts}
\usepackage{epsfig}
\usepackage[font={small}]{caption}

\usepackage[letterpaper,
left=.75in, right=.75in,
top=1in, bottom=1in
]{geometry}

\usepackage{psfrag}	
\usepackage{etoolbox}
\usepackage{algorithmicx}
\usepackage[Algorithm,ruled]{algorithm}
\usepackage{algpseudocode}
\usepackage{pifont}
\usepackage[utf8]{inputenc}
\usepackage[T1]{fontenc}  
\usepackage[nolist]{acronym}
\MakePerPage{footnote}
\usepackage{paralist}
\usepackage{enumitem}
\usepackage{bbm}
\usepackage[process=auto]{pstool}
\usepackage{tikz,pgfplots}
\usetikzlibrary{shapes,arrows}

\title{A Single-RF Architecture for Multiuser Massive MIMO via Reflecting Surfaces}
\name{Ali Bereyhi$^\star$,
Vahid Jamali$^\star$,
Ralf R. M\"uller$^\star$,
Antonia M. Tulino$^\dagger$,
Georg Fischer$^\star$, and
Robert Schober$^\star$\thanks{Emails: \{\textit{ali.bereyhi, vahid.jamali, ralf.r.mueller, georg.fischer, robert. schober}\}\textit{@fau.de}, and \textit{a.tulino@nokia-bell-labs.com}
}
\thanks{This work has been accepted for presentation in the 45th IEEE International Conference on Acoustics, Speech, and Signal Processing (ICASSP) 2020. The link to the final version in the Proceedings of ICASSP will be available later.}
}
\address{
$^\star$Friedrich-Alexander Universit\"at Erlangen-N\"urnberg\\
$^\dagger$Nokia Bell Labs, and Universita degli Studi di Napoli Federico II
}
\include{commands}

\begin{document}
\ninept
\maketitle

\begin{acronym}
\acro{mimo}[MIMO]{multiple-input multiple-output}
\acro{csi}[CSI]{channel state information}
\acro{awgn}[AWGN]{additive white Gaussian noise}
\acro{iid}[i.i.d.]{independent and identically distributed}
\acro{uts}[UTs]{user terminals}
\acro{bs}[BS]{base station}
\acro{tas}[TAS]{transmit antenna selection}
\acro{glse}[GLSE]{generalized least square error}
\acro{rhs}[r.h.s.]{right hand side}
\acro{lhs}[l.h.s.]{left hand side}
\acro{wrt}[w.r.t.]{with respect to}
\acro{mmW}[mmWave]{millimeter-wave}
\acro{np}[NP]{non-deterministic polynomial-time}
\acro{papr}[PAPR]{peak-to-average power ratio}
\acro{rzf}[RZF]{regularized zero forcing}
\acro{snr}[SNR]{signal-to-noise ratio}
\acro{rf}[RF]{radio frequency}
\acro{rfc}[RFC]{RF chain}
\acro{mf}[MF]{matched filtering}
\acro{gamp}[GAMP]{generalized AMP}
\acro{amp}[AMP]{approximate message passing}
\acro{vamp}[VAMP]{vector AMP}
\acro{map}[MAP]{maximum-a-posterior}
\acro{mmse}[MMSE]{minimum mean squared error}
\acro{mse}[MSE]{mean squared error}
\acro{ap}[AP]{average power}
\acro{pa}[PA]{power amplifier}
\acro{tdd}[TDD]{time division duplexing}
\acro{rss}[RSS]{residual sum of squares}
\acro{rls}[LS]{least-squares}
\acro{dbb}[DBB]{digital base-band}
\acro{ra}[RA]{reflect-array}
\acro{ta}[TA]{transmit-array}
\acro{had}[HAD]{hybrid analog-digital}
\acro{kkt}[KKT]{Karush-Kuhn-Tucker}
\end{acronym}

\begin{abstract}
In this work, we propose a new single-RF MIMO architecture which enjoys high scalability and energy-efficiency. The transmitter in this proposal consists of a single RF illuminator radiating towards a reflecting surface. Each element on the reflecting surface re-transmits its received signal after applying a phase-shift, such that a desired beamforming pattern is obtained. For this architecture, the problem of beamforming is interpreted as linear regression and a solution is derived via the method of least-squares. Using this formulation, a fast iterative algorithm for tuning of the reflecting surface is developed. Numerical results demonstrate that the proposed architecture is fully compatible with current designs of reflecting surfaces.
\end{abstract}



\section{Introduction}
\label{sec:intro}
A single-\ac{rf} transmitter consists of a monotone signal generator oscillating at the carrier frequecy and a network of tunable elements \cite{kalis2008novel,mohammadi2012single,sedaghat2014novel,sedaghat2016load}. These elements shape various output signals, depending on the value of their tunable factors. Using~this~architecture, downlink precoding is addressed completely in the \ac{rf} stage. Due to their high cost-efficiency, these architectures have received a great deal of attention in the context of massive \ac{mimo} systems. 

The conventional single-\ac{rf} architectures often employ elements such as \textit{parasitic antenna arrays} or \textit{load modulators} \cite{mohammadi2012single,sedaghat2016load}. Such settings require a physical module which connects the \ac{rf} signal generator to the analogue network. This introduces undesired \textit{power-loss} to the system resulting in performance degradation. An efficient way to address this issue is to implement the network of tunable elements via a \textit{reflecting surface}. The connection between the \ac{rfc} and the analog front-end in this case is carried out over the air which can lead to high power-efficiency and scalability \cite{jamali2019scale}. 

Inspired by promising performance gains obtained via reflecting surfaces, several recent studies have proposed reflecting-surface-assisted architectures for \ac{mimo} transmission. In \cite{basar2019transmission}, reflecting surfaces have been employed to improve signal transmission in a point-to-point communication scenario. This idea was further extended to multiuser scenarios in \cite{liu2019symbol}, where a symbol-level precoding scheme is designed via intelligent reflecting surfaces for downlink transmission in multiuser \ac{mimo} settings. Some recent studies in this respect can be followed in \cite{jamali2019scale,basar2019transmission,badiu2019communication,liu2019symbol,wu2019intelligent,huang2019reconfigurable} and the references therein.

\subsection{Contributions}
This paper proposes a novel reflecting-surface-assisted single-\ac{rf} \ac{mimo} architecture. In this architecture, the modulation and beamforming are carried out at the reflecting surface via the phase-shifts which are applied by reflecting elements. Unlike available proposals, e.g., \cite{liu2019symbol}, this architecture does not impose any restriction on the system and considers a \textit{generic} downlink transmission scenario. Invoking the \ac{rls} method, a tractable algorithm for tuning the \ac{rfc} and reflecting surface is developed. The investigations demonstrate that the proposed transmitter performs promisingly and is compatible with the available technology.


\subsection{Notations}
Throughout this paper, scalars, vectors, and matrices are represented by non-bold, bold lower case, and bold upper case letters, respectively. $\mH^{\her}$ and $\mH^{\sf P}$ denote the conjugate transpose and pseudo inverse of $\mH$, respectively. $\mI_K$ represents a $K\times K$ identity matrix. $\rho_{\max}^2\brc{\mH}$ returns the maximum squared singular value of $\mH$. $\norm{\cdot}$ and $\norm{\cdot}_F$ denote the Euclidean and Frobenius norm, respectively. $\setR$ is the real axis, and $\setC$ represents the complex plane. $\angle s$ denotes the phase of complex scalar $s$. $\Ex{\cdot}{}$ is mathematical expectation. For simplicity, $\set{1, \ldots , N}$ is abbreviated by $[N]$. 

\section{Problem Formulation}
\label{sec:sys}
Consider a multiuser setting in which a \ac{bs} intends to transmit messages to $K$ single-antenna users. The \ac{bs} is equipped with a \textit{single} transmit \ac{rfc} and a reflecting surface with $M$ antenna elements. It is assumed that the \ac{bs} knows the \ac{csi} of the downlink channels from the surface to the users.

The \ac{bs} pursues the following steps for transmission: First,~it~encodes the message of user $k$ to a codeword of length $N$, i.e., $s_k\brc{1},$ $\ldots, s_k\brc{N}$ for $k\in\dbc{K}$. It then constructs transmit signal $\bx\brc{n}\in$ $\setC^M$ for $n\in\dbc{N}$ from the \ac{csi} and vector of information symbols $\bs\brc{n} \coloneqq \dbc{ s_1\brc{n},\ldots,s_K \brc{n} }^\trp$ using a \textit{single-RF transmitter}.

Following the above steps, $\bx\brc{1},\ldots,\bx \brc{N}$~are~transmitted via the reflecting surface to the users in $N$ distinct transmission time intervals. Let channel matrix $\mH \in \setC^{K\times M}$ contain the downlink channel coefficients. We assume that the channel experiences \textit{semi-static} fading meaning that $\mH$ remains unchanged during the transmission. Consequently, the vector of received signals in interval $n$ reads
\begin{align}
\by\brc{n} = \mH \left. \bx\brc{n} \right. + \bz\brc{n},
\end{align}
where $\bz\brc{n}$ is \ac{awgn} with zero mean and variance $\sigma^2$, i.e., sequence $\set{\bz\brc{n}: n\in\dbc{N}}$ is \ac{iid} and $\bz\brc{n} \sim \mathcal{CN}\brc{\boldsymbol{0}, \sigma^2 \mI_K}$ for each $n\in\dbc{N}$. $\by\brc{n}$ further reads $\by\brc{n} = \dbc{ y_1\brc{n},\ldots,y_K \brc{n} }^\trp$ with $y_k\brc{n}$ representing the signal received by user $k$ in interval $n$.

\subsection{Transmitter Architecture}
\label{sec:Reflect}
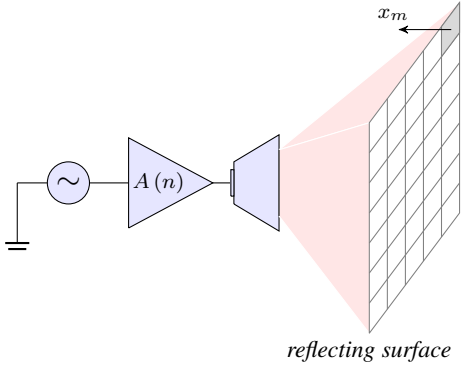
\begin{figure}[t]
\centering
\input{Figs/SysModel.tex}
\caption{Reflecting-surface-assisted single-RF MIMO transmitter.}
\label{fig:Sys}
\end{figure}

The detailed architecture of the transmitter is shown in Fig.~\ref{fig:Sys}. In this architecture, the \ac{rfc} is fed with a monotone signal which oscillates at the carrier frequency. The signal is first \textit{amplified} and then \textit{radiated} towards a reflecting surface whose center is located in distance $R_{\rm d}$ from the \ac{rfc}. The elements on this surface receive \textit{attenuated} and \textit{phase-shifted} versions of the radiated signal. Each element reflects its received signal after applying a \textit{tunable} phase-shift.

Let $P$ be the power of the monotone signal and $A\brc{n}$ denote the amplification gain of the \ac{rfc} in the $n$-th transmission time interval. For $m\in\dbc{M}$, the signal transmitted via the $m$-th antenna element on the reflecting surface is given by
\begin{align}
x_m \brc{n} = A\brc{n} T_m \sqrt{P} \exp\set{\rmj \dbc{ \omega_m + \beta_m\brc{n} } }
\end{align}
where $T_m$ and $\omega_m$ denote the attenuation and the phase-shift imposed on the radiated signal at the place of antenna element $m$, due to propagation, respectively. $\beta_m \brc{n}$ is  moreover the phase-shift~applied by the $m$-th antenna element before reflection. Depending on the properties of the surface, $\beta_m\brc{n}$ is chosen from $\setB \subseteq \dbc{-\pi , \pi}$. 

Fig.~\ref{fig:Sys} represents a special form of the architecture studied in \cite{jamali2019scale,bereyhi2019papr} in which the digital base-band unit is omitted and all the signal processing tasks are shifted to the analog unit. Following derivations in \cite{jamali2019scale}, it is concluded that
\begin{subequations}
\begin{align}
T_m &= \frac{\lambda \sqrt{ \zeta G\brc{\theta_{m} , \phi_{m}}  }}{4 \pi r_{m } }, \\
\omega_m &= - \frac{2\pi r_{m} }{\lambda},
\end{align}
\end{subequations}
where $\lambda$ is the wavelength, $\zeta$ denotes the power efficiency of the antenna elements, and $\brc{r_{m}, \theta_{m}, \phi_{m}}$ is the spherical coordinate of the $m$-th element on the surface when the origin is set at the \ac{rfc}. $G\brc{\theta , \phi}$ further represents the radiation pattern of the \ac{rfc} with $\theta$ and $\phi$ being the elevation and azimuth angle, respectively.

\section{LS-Based Single-RF Transmitter}
In the proposed reflecting-surface-assisted architecture, the signal pre-processing tasks are carried out via the tunable phase-shifts and amplitude variations at the \ac{rfc}. Hence, the phase-shifters and amplification gain are updated in each transmission time interval based on the given information symbols. As a result, the design reduces to the problem of finding $\beta_m \brc{n}$ for $m\in\dbc{M}$ and $A\brc{n}$ in terms of $\bs\brc{n}$ and $\mH$, such that~\textit{a desired performance metric} is optimized. We address~this problem via the method of \ac{rls}.

\subsection{Transmission with Minimum Received MSE}
The ultimate goal of transmitter design is to construct $\bx\brc{n}$, such that user $k$ can recover its information symbol $s_k\brc{n}$ with minimal post-processing from the received symbol $y_k\brc{n}$. To quantify this requirement, let us define the \textit{received \ac{mse}} in this setting as
\begin{align}
\mseR \brc{n} \coloneqq \sum_{k=1}^K \Ex{ \abs{ s_k\brc{n} - G_k y_k\brc{n}}^2 \left\vert \bs\brc{n} , \mH \right. }{} \label{eq:MSER}
\end{align}
for some $G_k\in\setC$. The received \ac{mse} determines the sum of \acp{mse} at the user terminals when user $k$ takes $y_k\brc{n}$ as the \textit{soft estimate} of its information symbols after \textit{amplification} and performing a \textit{phase-shift}. In \eqref{eq:MSER}, $G_k$ models linear~post-processing operations on $y_k\brc{n}$, i.e., amplification and phase-shift, and is assumed to be independent of $n$. The expectation is further calculated conditioned on $\bs\brc{n}$ and $\mH$, since the information symbols and the \ac{csi} are known at the transmitter side.

Using the independence of \ac{awgn}, it is straightforward to show that the received \ac{mse} reduces to
\begin{align}
\mseR \brc{n} \coloneqq \norm{ \bs\brc{n} - \mG \mH \bx\brc{n}}^2 + \norm{\mG}_F^2 \sigma^2 \label{eq:MSER2}
\end{align}
with $\mG = \diag{G_1,\ldots,G_K}$. From the transmitter architecture, we further have
\begin{align}
\bx\brc{n} = A\brc{n} \sqrt{P} \mT \bw\brc{n},
\end{align}
where $\mT$ and $\bw\brc{n}$ are given by
\begin{subequations}
\begin{align}
&\mT = \diag{T_1\exp\set{\rmj \omega_1},\ldots,T_M\exp\set{\rmj \omega_M}}.\\
&\bw\brc{n} = \dbc{\exp\set{\rmj \beta_1\brc{n}},\ldots,\exp\set{\rmj \beta_M\brc{n}}}^\trp.
\end{align}
\end{subequations}
By substituting $\bx\brc{n}$ in \eqref{eq:MSER2}, the received \ac{mse} reads
\begin{align}
\mseR \brc{n} \hspace*{-.5mm} \coloneqq \hspace*{-.5mm} \norm{ \bs\brc{n} \hspace*{-.5mm} - \hspace*{-.5mm} A\brc{n} \sqrt{P} \mG \mH \mT \bw\brc{n}}^2 + \norm{\mG}_F^2 \sigma^2. \label{eq:MSER3}
\end{align}

Following the intuitive discussion at the beginning of the section, the design goal can be formulated via an optimization problem which minimizes the received \ac{mse}. This means that the optimal choices for the phase-shifts and amplification gain in transmission time interval $n$ are given by
\begin{align}
\dbc{\bw^\star\brc{n},A^\star\brc{n}} = \argmin_{ \bw\in \setW^M , A\in\setR } \norm{ \bs\brc{n} - \left. A \right. \sqrt{P}  \mG \mH \mT \bw }^2, \label{eq:Precoder_MMSE}
\end{align}
where $\setW$ is defined as
\begin{align}
\setW = \set{ \rmw = \exp\set{\rmj \beta }: \beta \in \setB }.
\end{align}
In other words, assuming that user $k$ multiplies its received signal with $G_k$ to calculate a soft estimate of its information symbol, the sum of \acp{mse} at the user terminals in time interval $n$ is minimized when we set $\beta_m\brc{n} = \angle \rmw_m^\star\brc{n}$ and $A\brc{n} = A^\star\brc{n}$.

The tuning scheme in \eqref{eq:Precoder_MMSE} can be observed as linear regression in which the regression coefficients in $\bw^\star\brc{n}$ is determined~via the \ac{rls} method from regressor matrix $A^\star\brc{n} \sqrt{P} \mG \mH \mT$ and the regressands in $\bs\brc{n}$. Such a problem in its basic form is trivially solved via \textit{zero-forcing}. However, the fact that the regression coefficients are restricted~to~lie on the unit circle makes the problem intractable.~To address this issue, we develop an iterative algorithm by modifying the \textit{gradient decent} algorithm.

\begin{remark}
\label{Rem:1}
In general, $\bw^\star\brc{n}$ and $A^\star\brc{n}$ for $n\in\dbc{N}$ are functions of the user processing gains. Considering this dependency, one can write $\mseR \brc{n} = f_n\brc{\mG}$ for some $f_n\brc{\cdot}$. This indicates that $G_k$, for $k\in\dbc{K}$, are optimally given by 
\begin{align}
\mG^\star  = \argmin_{ \mG \in \diag{\setC^K}} \frac{1}{N}\sum_{n=1}^N f_n \brc{\mG}.
\label{eq:OptG}
\end{align}
This task is notionally different from the tuning task of the transmit phase-shifters whose update-rate is much faster. Noting that the explicit form of $f_n\brc{\cdot}$ is not known, this optimization is intractable to be addressed. One can however employ an \textit{alternating optimization} approach to tune $\mG$ suboptimally. 
\end{remark}

\subsection{Developing an Iterative Algorithm}
When $\setB = \dbc{-\pi,\pi}$ and $A = 1$, the optimization problem in \eqref{eq:Precoder_MMSE} reduces to the so-called \textit{unit-modulus \ac{rls}}; see \cite{lu2013novel,Soltanalian2014Design,tranter2017fast} and the references therein. The benchmark approach for solving unit-modulus \ac{rls} is to apply semidefinite relaxation \cite{luo210sdr}. The computational complexity in this case grows quadratically in $M$ which is relatively large for the given application. In \cite{tranter2017fast}, a fast iterative algorithm for unit-modulus \ac{rls} based on gradient projection was proposed. The algorithm employs gradient decent approach to find a local minima. To keep the minima on the unit circle, it projects the updated point in each iteration on the unit circle. It was shown in \cite{tranter2017fast} that this algorithm globally converges to a \ac{kkt} point of the~unit-modulus \ac{rls} problem. The algorithm was further extended to \textit{auto-scaled} unit-modulus \ac{rls}, i.e., when the regressands are scaled with a tunable variable.

The key difference of \eqref{eq:Precoder_MMSE} to an auto-scaled unit-modulus~\ac{rls}~problem is that the support of regression coefficients, i.e., $\setW$, is not necessarily the unit circle. Following the gradient projection approach in \cite{tranter2017fast}, this issue can be addressed by straightforward modifications. The result is shown in Algorithm~\ref{GP}. In this algorithm, $\mathrm{Quant_{\setW}}\brc{\cdot}$ denotes a \textit{uniform} phase quantizer which for $\buu\in\setC^M$ is defined as
\begin{align}
\dbc{\mathrm{Quant_{\setW}}\brc{\buu}}_m = \argmin_{\rmw \in \setW} \left. \abs{ \angle \rmw - \angle \rmu_m } \right. .
\end{align}
By setting $\setW$ to be the unit circle in the complex plane, this algorithm recovers the gradient projection algorithm given in \cite{tranter2017fast}. 

\begin{algorithm}[t]
\caption{Gradient Projection Algorithm}
\label{GP}
\begin{algorithmic}[0]
\Initiate Set initial step-size $\psi_0\in\brc{0,1}$, and $\tilde{\mH} = \sqrt{P} \mG \mH \mT$
\begin{align*}
\bw_0 \brc{n} = \mathrm{Quant_{\setW}}\brc{\frac{\tilde{\mH}^{\sf P} \bs\brc{n} }{\abs{\tilde{\mH}^{\sf P} \bs\brc{n}}} }
\end{align*}
Choose $E_{\rm Th}$ and maximum number of iterations $T_{\max}$.\vspace*{1mm}
\While\NoDo $\norm{\bw_{t+1}-\bw_t}^2 \geq E_{\rm Th}$ and $t \leq T_{\max}$\vspace*{.5mm}
\begin{itemize}
\item[{$\blacktriangleright$}] Update the parameters of the \ac{rfc} as
\begin{align*}
A_{t+1} \brc{n} =  \frac{ \Re \set{ \bw_t^\her\brc{n} \tilde{\mH}^{\her} \bs\brc{n} } }{  \norm{\tilde{\mH} \bw_t}^2 }.
\end{align*}
\item[{$\blacktriangleright$}] Update the step-size as
\begin{align*}
\psi_{t+1} = \frac{\psi_0 A_{t+1} \brc{n} }{ \rho_{\max}^2\brc{A_{t+1} \brc{n} \tilde{\mH}} }
\end{align*}
\item[{$\blacktriangleright$}] Update the parameters of the surface as
\begin{align*}
\bvv_{t+1} \brc{n} &= \tilde{\mH}^{\her} \brc{\bs\brc{n} - A_{t+1} \brc{n} \tilde{\mH} \bw_t\brc{n} }\\
\bw_{t+1} \brc{n} &= \mathrm{Quant_{\setW}}\brc{\bw_t\brc{n} + \psi_{t+1}  \bvv_{t+1}\brc{n}}
\end{align*}
\item[{$\blacktriangleright$}] Update $t \leftarrow t+1$.
\end{itemize}
\EndWhile \vspace*{1mm}
\Out $A\brc{n} = A_{T} \brc{n}$ and $\beta_m\brc{n} = \angle{\rmw_{T,m} \brc{n}}$ for $m\in\dbc{M}$, where $T$ is index of the last iteration.
\end{algorithmic}
\end{algorithm}

\section{Numerical Experiments}
We now investigate performance of the proposed architecture~through some numerical experiments. To this end, we consider the following scenario: The \ac{rfc} is fed by an oscillator with power $P=1$ and wavelength $\lambda =8$ mm. The transmit antenna at the \ac{rfc} has a horizontally omnidirectional radiation pattern whose beamwidth in the vertical plane is $120^\circ$. A reflecting surface of size $\sqrt{M} \lambda \times \sqrt{M} \lambda$ with $M$ elements is located in distance $R_{\rm d} = \lambda \sqrt{M/\pi}  $ from the \ac{rfc}. The power efficiency of the elements is set to $\log \zeta = 0$ dB and the phase-shifts are quantized with $B$ bits, i.e.,
\begin{align}
\setB = \set{ -\pi + \frac{ \left. i \right. \pi}{2^{B-1}}  \ \text{ for } \ i=0,\ldots,2^B-1 }.
\end{align}

The architecture is used for downlink transmission to $K$ users in a single cell. The users are assumed to be \textit{uniformly} distributed in the network. We consider the standard~Rayleigh~model for the fading process and model the shadowing effects via the log-normal distribution. The entries of $\mH$ are hence generated~as
\begin{align}
\dbc{\mH}_{k,m} = \left. \sqrt{\frac{\alpha_k}{\bar{\mathrm{r}}_k^\nu} } \right. h_{k,m},
\end{align}
where 
\begin{itemize}
\item $h_{k,m}$ for $k\in\dbc{K}$ and $m\in\dbc{M}$ are \ac{iid} complex Gaussian random variables with zero mean and unit variance,
\item $\alpha_k$ for $k\in\dbc{K}$ are zero-mean log-normal variables with standard deviation $\log \sigma_{\rm Shadow} = 5$ dB,
\item  $\nu = 3.2$ is the attenuation exponent, and
\item  $\bar{\mathrm{r}}_k = {\mathrm{r}}_k / {\mathrm{r}}_{\rm h}$ with ${\mathrm{r}}_k$ being the distance of user $k$ to the \ac{bs} and ${\mathrm{r}}_{\rm h} = 100$ m denoting the minimal distance in the network.
\end{itemize}

Throughout numerical simulations, we consider encoded messages of length $N=100$. The information symbols in each transmission interval are assumed to be \ac{iid} zero-mean and unit-variance complex Gaussian random variables, i.e., $\bs\brc{n} \sim \mathcal{CN}\brc{\boldsymbol{0}, \mI_K}$. The post-processing gains at user terminals are further set such that path-loss and shadowing effects are compensated. Note that based on the discussion in Remark~\ref{Rem:1}, this choice of $\mG$ is in general suboptimal.

To evaluate the performance of this setting, we define the \textit{per-user distortion} as
\begin{align}
D =  \frac{1}{K} \brc{ \frac{1}{N} \sum_{n=1}^N \norm{ \bs\brc{n} - \mG \mH \bx \brc{n} }^2 }
\end{align}
for a given realization of the channel. For this setting,~the~\textit{average transmit power} per time interval is further calculated as
\begin{align}
P_{\rm Out} =   \frac{1}{N} \sum_{n=1}^N  \abs{A\brc{n}}^2 P.
\end{align}
Noting that the \ac{rfc} transmits with power $A^2\brc{n} P$ in interval $n$, we define the \textit{\ac{papr}}~as 
\begin{align}
\papr =   \frac{\max_{n\in\dbc{N}}  \abs{A\brc{n}}^2 P}{ P_{\rm Out} }.
\end{align}
These parameters are averaged over multiple channel realizations.

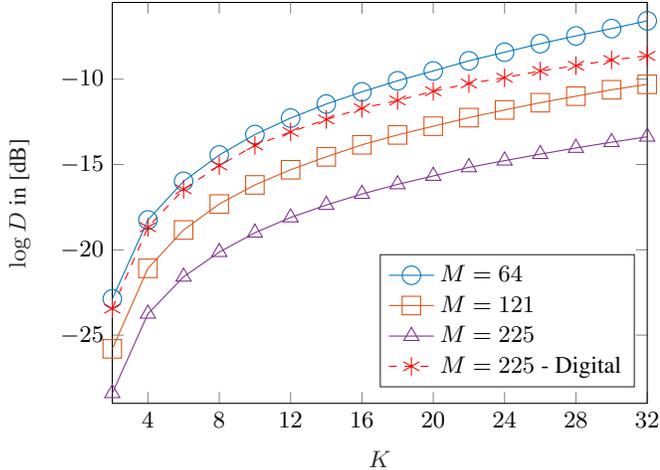
\begin{figure}
\centering
\input{Figs/Fig1.tex}
\caption{Per-user distortion vs. $K$ for various scenarios.}
\label{fig:1}
\end{figure}

Fig.~\ref{fig:1} shows the per-user distortion against the number of user for various values of $M$. Here, the phase-shifts are quantized with $B=4$ bits. From the figure, it is observed that $D$ increases, as $K$ grows. This indicates that the distortion scales reversely with the number of antenna elements per user terminal. The variations in $D$ with respect to the surface size further agrees with this conclusion. To compare the performance with a benchmark, we further plot the curve for a fully digital matched filtering considering the case with $M = 225$ antennas. In this scheme, a fully digital transmitter, i.e., a transmitter with $M = 225$ distinct \acp{rfc}, uses matched filtering for downlink transmission. In each transmit interval, the digital precoder is scaled, such that its transmit power is exactly the same as the one radiated in the proposed single-\ac{rf} architecture. User $k$ further chose $G_k$, such that the impact of path-loss, shadowing and power scaling on its information symbol is compensated. From the figure, it is observed that the proposed architecture outperforms this fully digital scheme with even with $M=121$ transmit antennas. Such an observation indicates the effectiveness of the proposed architecture with respect to the benchmark.

\begin{figure}
\centering
\input{Figs/Fig2.tex}
\caption{PAPR vs. $K$ for various sizes of the reflecting surface.}
\label{fig:2}
\end{figure}
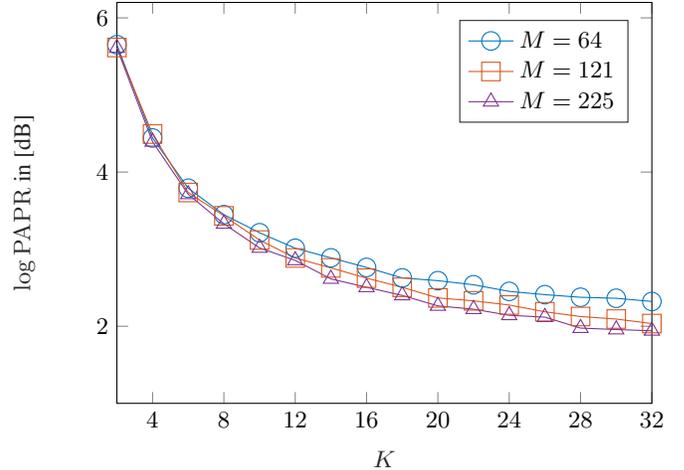

Although the growth in $K$ increases the per-user distortion, it can benefit in terms of \ac{papr} of the transmit \ac{rfc}. This is demonstrated in Fig. \ref{fig:2}, where the transmit \ac{papr} is plotted for the same settings against $K$. This behavior comes from the fact that as the number of users increases, the transmit signal couples higher number of information symbols which by the law of large numbers reduces the time variations.

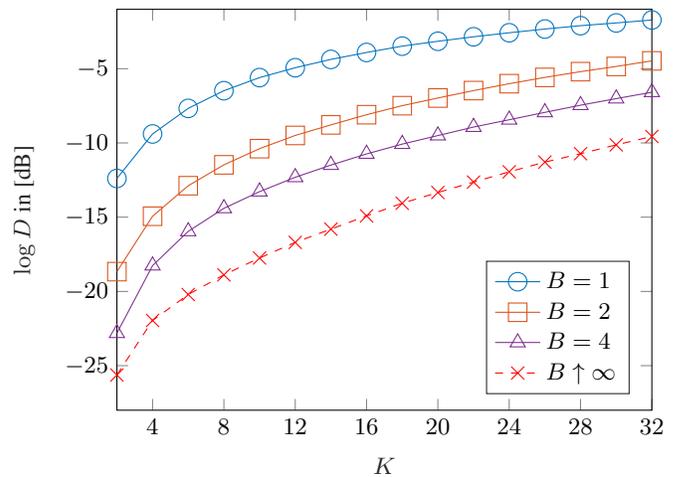
\begin{figure}
\centering
\input{Figs/Fig3.tex}
\caption{Impact of phase-shift quantization at the reflecting surface.}
\label{fig:3}
\end{figure}

Fig.~\ref{fig:3} illustrates the impact of phase quantization. Here, distortion is plotted against $K$ for various choices of $B$ when $M=64$. As the figure demonstrates, by increasing the resolution of the phase-shifters, the performance improves. Comparing the results with the asymptotic case of $B\to \infty$, i.e., when $\setB = \dbc{-\pi,\pi}$, it is observed that the degradation due to the phase-shift quantization is not significant. This indicates that the proposed architecture is compatible with the available designs for reflecting surfaces \cite{chou2018all,abdelrahman2017analysis}.

\section{Conclusions}
A reflecting-surface-assisted single-\ac{rf} \ac{mimo} architecture has been proposed. The transmitter consists of an illuminator~and a passive reflecting surface. This makes the architecture both cost-efficient and scalable. To tune the phase-shifters at the surface, a fast algorithm based on the method of \ac{rls} has been developed. Simulations have demonstrated that desired per-user distortions are achievable via this architecture using available technologies of reflecting surfaces. The study in this work has considered a basic form of the proposed architecture and requires further investigations and extensions in various respects. The work in this direction is currently~ongoing.

\bibliographystyle{IEEEbib}
\bibliography{ref}

\end{document}

%% file: commands.tex
\newcommand{\setR}{\mathbbmss{R}}

\newcommand{\setW}{\mathbbmss{W}}

\newcommand{\setB}{\mathbbmss{B}}

\newcommand{\setC}{\mathbbmss{C}}

\newcommand{\Ex}[2]{ \mathbbm{E}_{#2} \left\lbrace #1 \right\rbrace }

\newcommand{\rmj}{\mathrm{j}}

\newcommand{\her}{\mathsf{H}}

\newcommand{\argmin}{\mathop{\mathrm{argmin}}}

\newcommand{\bw}{\mathbf{w}}
\newcommand{\buu}{\mathbf{u}}
\newcommand{\bvv}{\mathbf{v}}

\newcommand{\rmw}{\mathrm{w}}
\newcommand{\rmu}{\mathrm{u}}

\newcommand{\papr}{\mathrm{PAPR}}

\newcommand{\bx}{{\boldsymbol{x}}}

\newcommand{\set}[1]{\left\lbrace#1\right\rbrace}
\newcommand{\diag}[1]{\mathrm{diag}\left\lbrace #1 \right\rbrace}
\newcommand{\brc}[1]{\left( #1 \right) }

\newcommand{\dbc}[1]{\left[ #1 \right] }

\newcommand{\bz}{{\boldsymbol{z}}}

\newcommand{\bs}{{\boldsymbol{s}}}

\newcommand{\by}{{\boldsymbol{y}}}

\newcommand{\trp}{\mathsf{T}}

\newcommand{\mI}{\mathbf{I}}

\newcommand{\mG}{\mathbf{G}}

\newcommand{\mT}{\mathbf{T}}
\newcommand{\mH}{\mathbf{H}}

\newcommand{\mseR}{\mathrm{MSE}_{ \mathsf{R} }}

\newcommand{\norm}[1]{\lVert #1 \rVert}

\newcommand{\abs}[1]{\lvert #1 \rvert}

\newtheoremstyle{mystyle}
  {}
  {}
  {}
  {}
  {\bfseries}
  {:}
  { }
  {}

\theoremstyle{mystyle}

\newtheorem{remark}{Remark}

\algnewcommand\algorithmicLet{\textbf{Let}}
\algnewcommand\Let{\item[\algorithmicLet]}
\algnewcommand\algorithmicSet{\textbf{Set}}
\algnewcommand\Set{\item[\algorithmicSet]}

\algnewcommand\algorithmicInitiate{\textbf{Initiate}}
\algnewcommand\Initiate{\item[\algorithmicInitiate]}
\algnewcommand\algorithmicStart{\textbf{Begin}}
\algnewcommand\Begin{\item[\algorithmicStart]}
\algnewcommand\algorithmicEnd{\textbf{End}}
\algnewcommand\End{\item[\algorithmicEnd]}

\algnewcommand\algorithmicOutP{\textbf{Output:}}
\algnewcommand\Out{\item[\algorithmicOutP]}
\newcommand\NoDo{\renewcommand\algorithmicdo{}}

\newcounter{bar}



%% file: Figs/SysModel.tex
\begin{tikzpicture}[scale=.8]

\tikzset{
    >=stealth',
    pil/.style={
           ->,
           thick,
           shorten <=2pt,
           shorten >=2pt,},
    pilak/.style={
           <->,
           thin,
           shorten <=2pt,
           shorten >=2pt,}
}

\draw (2.5,-2.8) node {\textit{reflecting surface}};



%

\draw (-3.5,0) node { };
\draw (-3.35,0) -- (-2.85,0);
\draw (-3.35,-1) -- (-3.35,0);
\draw[thick] (-3.55,-1) -- (-3.15,-1);
\draw[thick] (-3.5,-1.1) -- (-3.2,-1.1);
\draw [fill=blue!10] (-2.5,0) circle (.35cm) node {\large{$\sim$}};
\draw (-2.15,0) -- (-1.5,0);
\filldraw[fill=blue!10] (-1.5,-.75) -- (-1.5,.75) -- (-.1,0) -- (-1.5,-.75);
\draw (-.1,0) -- (.2,0);
\filldraw[fill=blue!10] (.25,-.35) -- (.25,.35) -- (1,.8) -- (1,-.8)-- (.25,-.35);
\filldraw[fill=blue!10] (.2,-.22) -- (.2,.22) -- (.25,.22) -- (.25,-.22)-- (.2,-.22);


\draw (-1,0) node[font=\footnotesize] {$A\brc{n}$};
%

\filldraw[red!10,white,fill=red!10] (1,.55) -- (2.5,1) -- (4,3) -- (1,.55);
\filldraw[red!10,white,fill=red!10] (1,.55) -- (2.5,1) -- (2.5,-2.5) -- (1,-.55)-- (1,.55);
\draw (1,-.55)-- (1,.55);

\draw[gray,fill=white] (2.5,-2.5) -- (2.5,1) --  (4,3) -- (4,-0.5) -- (2.5,-2.5);

\draw[gray] (2.5,-2) -- (4,0);
\draw[gray] (2.5,-1.5) -- (4,0.5);
\draw[gray] (2.5,-1) -- (4,1);
\draw[gray] (2.5,-.5) -- (4,1.5);
\draw[gray] (2.5,0) -- (4,2);
\draw[gray] (2.5,.5) -- (4,2.5);

\draw[gray] (2.8,1.4) -- (2.8,-2.1);
\draw[gray] (3.1,1.8) -- (3.1,-1.7);
\draw[gray] (3.4,2.2) -- (3.4,-1.3);
\draw[gray] (3.7,2.6) -- (3.7,-0.9);
\filldraw[gray,fill=gray!30] (4,3) -- (3.7,2.6) --  (3.7,2.1) -- (4,2.5) -- (4,3);


\draw[pilak,->] (3.95,2.55) -- (2.9,2.55);
\draw (2.9,2.8) node[font=\footnotesize] {$x_m$};

\end{tikzpicture}

%% file: Figs/Fig1.tex
%
%
\definecolor{mycolor1}{rgb}{0.00000,0.44700,0.74100}%
\definecolor{mycolor2}{rgb}{0.85000,0.32500,0.09800}%
\definecolor{mycolor3}{rgb}{0.49400,0.18400,0.55600}%
\begin{tikzpicture}

\begin{axis}[%
width=2.8in,
height=2.1in,
at={(1.962in,0.944in)},
scale only axis,
xmin=2,
xmax=32,
xtick={4,8,12,16,20,24,28,32},
xticklabels={{$4$},{$8$},{$12$},{$16$},{$20$},{$24$},{$28$},{$32$}},
xlabel style={font=\color{white!15!black}},
xlabel={$K$},
ymin=-29,
ymax=-5.5,
ytick={-25,-20,-15,-10,-5},
yticklabels={{$-25$},{$-20$},{$-15$},{$-10$},{$-5$}},
ylabel style={font=\color{white!15!black}},
ylabel={$\log D$ in [dB]},
axis background/.style={fill=white},
legend style={at={(0.5,0.04)}, anchor=south west, legend cell align=left, align=left, draw=white!15!black}
]
\addplot [color=mycolor1, mark size=3.5pt, mark=o, mark options={solid, mycolor1}]
  table[row sep=crcr]{%
2	-22.8782846446457\\
4	-18.2504897515909\\
6	-15.982967790644\\
8	-14.4428783478083\\
10	-13.2454118111701\\
12	-12.2788190731767\\
14	-11.4570339377232\\
16	-10.7397419517572\\
18	-10.0945516667657\\
20	-9.51555126965926\\
22	-8.92193614540693\\
24	-8.41659067165364\\
26	-7.91654204158618\\
28	-7.46361424846525\\
30	-7.0354274450017\\
32	-6.57977731002221\\
};
\addlegendentry{$M=64$}

\addplot [color=mycolor2, mark size=3.5pt, mark=square, mark options={solid, mycolor2}]
  table[row sep=crcr]{%
2	-25.8049863854436\\
4	-21.0888484156603\\
6	-18.8383312213934\\
8	-17.3148160227384\\
10	-16.1951717954961\\
12	-15.3064997597411\\
14	-14.5507148761672\\
16	-13.8480931861076\\
18	-13.2663067768094\\
20	-12.7566266818072\\
22	-12.2540420415794\\
24	-11.8029221498484\\
26	-11.3781112966971\\
28	-10.9954072185624\\
30	-10.6201221184899\\
32	-10.3005153822789\\
};
\addlegendentry{$M=121$}

\addplot [color=mycolor3, mark size=3.3pt, mark=triangle, mark options={solid, mycolor3}]
  table[row sep=crcr]{%
2	-28.4089206437786\\
4	-23.7504314521827\\
6	-21.5814365874326\\
8	-20.1359912569445\\
10	-19.0002522856868\\
12	-18.10684079776\\
14	-17.3749123583638\\
16	-16.7291812643092\\
18	-16.1654562997952\\
20	-15.6708140253133\\
22	-15.1675640113823\\
24	-14.7749147220169\\
26	-14.3942089103687\\
28	-14.0315608726524\\
30	-13.693023675769\\
32	-13.3758374927585\\
};
\addlegendentry{$M=225$}

\addplot [color=red, dashed, mark size=3.5pt, mark=asterisk, mark options={solid, red}]
  table[row sep=crcr]{%
2	-23.4469706205184\\
4	-18.6788844020144\\
6	-16.4536200802185\\
8	-15.0624704694576\\
10	-13.8819756823598\\
12	-13.0913055957646\\
14	-12.3628436148756\\
16	-11.704743240055\\
18	-11.2483096037141\\
20	-10.7076378559346\\
22	-10.2524230815653\\
24	-9.90963887101024\\
26	-9.51894389980676\\
28	-9.20824888543096\\
30	-8.87247670699127\\
32	-8.63141312535093\\
};
\addlegendentry{$M=225$ - Digital}

\end{axis}
\end{tikzpicture}%

%% file: Figs/Fig2.tex
%
%
\definecolor{mycolor1}{rgb}{0.00000,0.44700,0.74100}%
\definecolor{mycolor2}{rgb}{0.85000,0.32500,0.09800}%
\definecolor{mycolor3}{rgb}{0.49400,0.18400,0.55600}%
\begin{tikzpicture}

\begin{axis}[%
width=2.8in,
height=2.1in,
at={(1.962in,0.944in)},
scale only axis,
xmin=2,
xmax=32,
xtick={4,8,12,16,20,24,28,32},
xticklabels={{$4$},{$8$},{$12$},{$16$},{$20$},{$24$},{$28$},{$32$}},
xlabel style={font=\color{white!15!black}},
xlabel={$K$},
ymin=1,
ymax=6.2,
ytick={2,4,6},
yticklabels={{$2$},{$4$},{$6$}},
ylabel style={font=\color{white!15!black}},
ylabel={$\log \papr$ in [dB]},
axis background/.style={fill=white},
legend style={at={(0.64,0.7)}, anchor=south west, legend cell align=left, align=left, draw=white!15!black}
]
\addplot [color=mycolor1, mark size=3.5pt, mark=o, mark options={solid, mycolor1}]
  table[row sep=crcr]{%
2	5.65228285959509\\
4	4.44182506009526\\
6	3.79012531262167\\
8	3.44732748339354\\
10	3.21256035717951\\
12	3.01174336227384\\
14	2.88709230928887\\
16	2.76304518377838\\
18	2.62684706225703\\
20	2.59098164088304\\
22	2.5379395551187\\
24	2.45206136459711\\
26	2.40944523115026\\
28	2.3757968403015\\
30	2.36266957297036\\
32	2.32131619491984\\
};
\addlegendentry{$M=64$}

\addplot [color=mycolor2, mark size=3.5pt, mark=square, mark options={solid, mycolor2}]
  table[row sep=crcr]{%
2	5.61545625869912\\
4	4.49278308382687\\
6	3.73475173978107\\
8	3.4321007224568\\
10	3.11824851702135\\
12	2.88565197489008\\
14	2.75702866854009\\
16	2.62382375226184\\
18	2.50513724067234\\
20	2.36714063870133\\
22	2.32957530632533\\
24	2.27516821178558\\
26	2.1876901587011\\
28	2.1261595009205\\
30	2.09337326287838\\
32	2.03476232477044\\
};
\addlegendentry{$M=121$}

\addplot [color=mycolor3, mark size=3.3pt, mark=triangle, mark options={solid, mycolor3}]
  table[row sep=crcr]{%
2	5.60941595553842\\
4	4.39238532279914\\
6	3.71094687777887\\
8	3.32798833782444\\
10	3.01591639579789\\
12	2.85437180768854\\
14	2.61395304836394\\
16	2.50541952251479\\
18	2.3996208160998\\
20	2.26377128739318\\
22	2.22115871067818\\
24	2.14190648215698\\
26	2.11816104542198\\
28	1.97592513609007\\
30	1.9574731758527\\
32	1.94113921676003\\
};
\addlegendentry{$M=225$}

\end{axis}
\end{tikzpicture}%

%% file: Figs/Fig3.tex
%
%
\definecolor{mycolor1}{rgb}{0.00000,0.44700,0.74100}%
\definecolor{mycolor2}{rgb}{0.85000,0.32500,0.09800}%
\definecolor{mycolor3}{rgb}{0.49412,0.18431,0.55686}%
\begin{tikzpicture}

\begin{axis}[%
width=2.8in,
height=2.1in,
at={(1.962in,0.944in)},
scale only axis,
xmin=2,
xmax=32,
xtick={4,8,12,16,20,24,28,32},
xticklabels={{$4$},{$8$},{$12$},{$16$},{$20$},{$24$},{$28$},{$32$}},
xlabel style={font=\color{white!15!black}},
xlabel={$K$},
ymin=-28,
ymax=-1,
ytick={-25,-20,-15,-10,-5},
yticklabels={{$-25$},{$-20$},{$-15$},{$-10$},{$-5$}},
ylabel style={font=\color{white!15!black}},
ylabel={$\log D$ in [dB]},
axis background/.style={fill=white},
legend style={at={(0.69,0.04)}, anchor=south west, legend cell align=left, align=left, draw=white!15!black}
]
\addplot [color=mycolor1, mark size=3.5pt, mark=o, mark options={solid, mycolor1}]
  table[row sep=crcr]{%
2	-12.3920904507959\\
4	-9.39583789775266\\
6	-7.67343194057881\\
8	-6.48217636555247\\
10	-5.60033101814594\\
12	-4.93592984498254\\
14	-4.37541572661711\\
16	-3.90746222246201\\
18	-3.4818624937081\\
20	-3.1526294739045\\
22	-2.85199861148535\\
24	-2.58001435928964\\
26	-2.3294559898067\\
28	-2.10301717101034\\
30	-1.91831762182791\\
32	-1.71982224249367\\
};
\addlegendentry{$B=1$}

\addplot [color=mycolor2, mark size=3.5pt, mark=square, mark options={solid, mycolor2}]
  table[row sep=crcr]{%
2	-18.6679330104503\\
4	-14.9370931265574\\
6	-12.8816986018582\\
8	-11.4747988673764\\
10	-10.3862339288387\\
12	-9.49568610190091\\
14	-8.78481448714189\\
16	-8.09905213638941\\
18	-7.49123271806094\\
20	-6.97437285644131\\
22	-6.4747963376221\\
24	-6.01084773043771\\
26	-5.58710471340128\\
28	-5.20004772895599\\
30	-4.84904823773655\\
32	-4.47383976376283\\
};
\addlegendentry{$B=2$}

\addplot [color=mycolor3, mark size=3.3pt, mark=triangle, mark options={solid, mycolor3}]
  table[row sep=crcr]{%
2	-22.8127421865936\\
4	-18.2618889851655\\
6	-15.9551927303362\\
8	-14.4048295160181\\
10	-13.2799068764538\\
12	-12.3140336190091\\
14	-11.4905697028372\\
16	-10.7399274195026\\
18	-10.0783855059617\\
20	-9.50330411430554\\
22	-8.93134036455465\\
24	-8.43828248499165\\
26	-7.9411174401018\\
28	-7.45892448263223\\
30	-7.01479597108285\\
32	-6.58592751750269\\
};
\addlegendentry{$B=4$}

\addplot [color=red, dashed, mark size=3.5pt, mark=x, mark options={solid, red}]
  table[row sep=crcr]{%
2	-25.6312580672737\\
4	-21.9538478273475\\
6	-20.207936002984\\
8	-18.885014875138\\
10	-17.7491078006511\\
12	-16.6977950522877\\
14	-15.8057041184784\\
16	-14.9116372618484\\
18	-14.0629723893406\\
20	-13.3370497446948\\
22	-12.6363872371841\\
24	-11.9436089600877\\
26	-11.2903043098792\\
28	-10.7252127813975\\
30	-10.1311599338118\\
32	-9.56503142500664\\
};
\addlegendentry{$B \uparrow \infty$}

\end{axis}
\end{tikzpicture}%